\documentclass{aastex}
\usepackage{emulateapj5}
\usepackage{times}

\def\RXTE{{\it RXTE}}
\def\Ginga{{\it Ginga}}

\def\tsr{\tau_{\rm sr}}
\def\tsd{\tau_{\rm sd}}
\def\thr{\tau_{\rm hr}}
\def\thd{\tau_{\rm hd}}
\def\keV{\,{\rm keV}}
\def\cm{\,{\rm cm}}

\shorttitle{TIME DOMAIN ANALYSIS OF VARIABILITY IN CYGNUS X-1}
\shortauthors{MACCARONE, COPPI, \& POUTANEN}

\begin{document}
\title{Time Domain Analysis of Variability in Cygnus X-1: \\
Constraints on the Emission Models}
\author{Thomas J. Maccarone,\altaffilmark{1} Paolo S. Coppi, \altaffilmark{1} 
and Juri Poutanen\altaffilmark{2}}

\altaffiltext{1}{Astronomy Department, Yale University, P.O.Box 208101, 
New Haven,CT 06520-8101, USA} 
\altaffiltext{2}{Stockholm Observatory, SE-133 36, Saltsj\"obaden, Sweden}

\date{}

%\label{firstpage}

\begin{abstract}
We  use  time domain  analysis  techniques  to  investigate the  rapid
variability  of  Cygnus  X-1.   We  show  that  the  cross-correlation
functions between hard  and soft energy bands reach  values very close
to unity  and peak at  a lag of  less than 2 millisecond  for energies
separated  by a factor  of 10.   This confirms  that the  process that
produces X-ray photons at  different energies is extremely coherent on
short time scales and  strongly constrains emission models proposed to
explain    Fourier-frequency-dependent   time   lags.     We   present
autocorrelation functions at different energies, and note their widths
decrease with  increasing energy.  We  show that the  extended Compton
corona model produces auto-correlation functions whose widths increase
with  increasing energy, that  the model  of cylindrical  waves moving
inward  through a  transition disk  has too  large a  peak lag  in the
cross-correlation  function.   Models   of  magnetic  flaring  and  of
drifting blobs in a hot corona can qualitatively fit the observations.

\end{abstract}


\section{Introduction}

The X-ray/$\gamma$-ray spectrum of an accreting black hole such as
Cygnus X-1 in its hard state can be represented as the sum of a few
components: a soft component associated with the emission from a cold
accretion disk, a hard tail extending up to a few hundred keV
associated with a hot ``corona'', and a Compton reflection bump
produced when hard X-rays are reflected from cold material in the
accretion disk (see, e.g., Zdziarski et al.  1997; Gierli\'nski et al.
1997; Poutanen 1998).  Spectral data suggest thermal Comptonization by
a medium with a temperature of about 100 keV as the origin of the hard
tail.  The observed X-ray spectral slopes and the amplitude of Compton
reflection can be used to determine the geometry of the system.
However, a variety of models fit the spectra (see Poutanen 1998;
Beloborodov 1999a; Zdziarski 2000), so the parameters of the accretion
flow are not be well constrained by spectral data alone.  Most of the
spectral models, however, are applied to time-averaged spectra despite
of the fact that the sources show rapid spectral variability (see,
e.g., Nolan et al.  1981; Negoro, Miyamoto, \& Kitamoto 1994; Feng,
Li, \& Chen 1999) implying rapid changes in the physical conditions.
Examining temporal characteristics should help break the degeneracy
among spectral models.

Time  domain  techniques  were used in the  early  days of X-ray  astronomy
before  the  statistical  samples  of data were  sufficient  to make use of
Fourier domain analyses (see, e.g.,  Weisskopf,  Kahn, \& Sutherland  1975;
Sutherland, Weisskopf, \& Kahn 1978; Priedhorsky et al.  1979; Nolan et al.
1981).  The asymmetry of the cross-correlation function (CCF) of Cygnus X-1
was discovered in $\sim 150$ sec observations by Priedhorsky et al.  (1979)
and  Nolan  et  al.  (1981).  They  showed  that  the  CCF  peaks  at a lag
$\lesssim  10-40$ ms.  Using data from {\it EXOSAT}, Page (1985)  confirmed
these  results  and  claimed  a $\sim 6$ ms  shift  of the  peak of the CCF
between the 5-14 keV and the 2-5 keV bands.  These are the last  papers, to
our knowledge, that present the CCFs despite  immense  advances in temporal
resolution,  photon  statistics, and duration of  observations.  Aside from
attempts to model  individual  shots (Lochner,  Swank, \& Szymkowiak  1991;
Negoro et al.  1994; Focke 1998; Feng et al.  1999), recent  analyses  have
concentrated on Fourier domain techniques.

The CCF asymmetry is related to the Fourier-frequency-dependent hard
time lags between different spectral bands discovered by \Ginga\
(Miyamoto et al.  1988).  These data gave new strong constraints on
spectral models.  However, in the Fourier domain it is difficult to
measure time lags at frequencies above $\sim 30$ Hz (see, e.g., Nowak
et al.  1999a) corresponding to the light travel time in a region of
the main energy dissipation.  On the other hand, in the time domain,
the CCFs can be measured accurately down to the lags as short as 2 ms.
Furthermore, time domain functions (containing, in principle, the same
information as their Fourier domain companions) highlight different
information which can be used to constrain emission models further.

We present the results of time domain analyses of Cygnus X-1 as observed by
the {\it Rossi X-Ray Timing Explorer} (\RXTE) and compare our observational
results with model predictions.

\begin{figure*}
%\bigskip
%%%%%%%%%%%%%%%%%%%%%%%%%%%%%%%%%%%%%%%%%%%%%
%\plotone{ccf_cygx1.ps}
\centerline{\epsfxsize=12cm \epsfbox{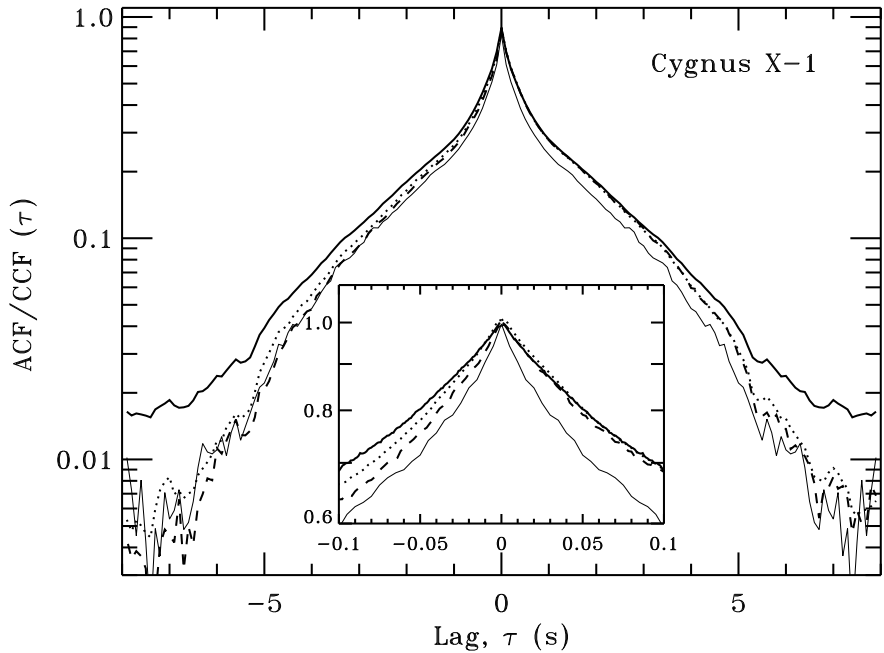}}
%\figcaption{
\caption{ The auto- and cross-correlation functions of Cygnus X-1
observed in December 1997.  Solid curves show the ACFs for the 2-5 keV
energy band (thick curves) and the 24-40 keV band (thin curves).
Dotted curves show the CCF between the 8-13 keV band and the 2-5 keV
band, and dashed curves represent the CCF for the 24-40 keV band vs
the 2-5 keV band.  The CCFs at other energies have similar shapes.
The peaks all align at around zero lag.  The higher energy curves are
narrower.  The CCFs are defined in such a way that the peak is
expected to appear at a positive lag when hard photons are lagging the
soft ones.  The errors at small lags are roughly the same size as the
line widths and are hence left unplotted for clarity.  The CCF peak
lag is at less than 2 ms for the 2-5 keV vs 8-13 keV CCF. 
%{\bf certainly not above a few seconds}
\label{fig:ccf}}
\end{figure*}

%\bigskip
%%%%%%%%%%%%%%%%%%%%%%%%%%%%%%%%%%%%%%%%%%%%%

\section{Observations and Results}

\subsection{Observation Log}

Cygnus X-1  was observed in  its hard (low)  state 12 times  by \RXTE\
during  December of  1997, for  a  total of  about 30  ksec.  Here  we
present results  only from the  Proportional Counter Array,  using the
standard   screening  criteria   of  earth   elevation   greater  than
$10\arcdeg$,  offset  from  source  less  than  $0.01\arcdeg$,  all  5
proportional counter  units on, and  the standard time since  the last
South  Atlantic Anomaly  passage.  For  the lowest  energies  (below 8
keV), we  have data  recorded in single  bit (SB) modes,  where \RXTE\
counts photons  with no spectral  information other than  whether they
fall  within the  given  range  of channels.   For  the higher  photon
energies,  we use  the full  Event Mode  data with  all  the available
spectral information.  The use of  SB modes and the subsequent loss of
some   spectral  information  were   required  because   of  telemetry
limitations of \RXTE.  We make no background subtractions for our data
because the background  counts are a very small  fraction of the total
counts and  because current background  models for \RXTE\ do  not give
estimates on time scales shorter than 16 seconds.  \RXTE\ deadtime for
Cygnus X-1 is about  1 \% and should affect only zero  lag bins in the
correlation functions.

\subsection{Light Curves}

Using the standard  FTOOLS 4.1 software, we extract  light curves with
2$^{-9}$ second resolution ($\sim$ 2 ms) for each of the SB modes (for
energy channels  2-5 keV, 5-6 keV,  and 6-8 keV), plus  the event mode
data binned into three additional energy bands (channels 24-40, 41-71,
and 72-132, or  8-13 keV, 13-24 keV, and  24-40 keV respectively).  We
do not  analyze any  higher energy data  because the  background count
rates become  large above  40 keV.  We  attempted to analyze  the data
with a $\sim  1 $ ms bin  size.  For the low energy  (and hence higher
count rate) bands the 1 ms computations allowed us to produce slightly
stronger constraints (discussed below).  We plot only the 2 ms binning
results  in order to  make the  plots clearer,  since the  results are
essentially the same in either case.

\subsection{Cross-correlation Functions}
\label{sect:ccf}

We compute  the CCF of Cygnus  X-1,  comparing  each  energy  band with the
lowest energy band.  The CCFs are asymmetric with peaks at lags less than 2
ms in all cases (see the inset of  Fig.~\ref{fig:ccf}),  and less than 1 ms
between  the 3 keV band and bands below 13 keV.  The rising part of the CCF
(soft lags)  becomes  narrower  with energy  substantially  faster than the
decaying part (hard lags).  These  results  agree  qualitatively  with past
results for the CCF  (Priedhorsky  et al.  1979;  Nolan et al.  1981).  The
CCFs reach values very close to unity,  showing  that the signal at all the
energies is extremely well  synchronized.  The asymmetry of the CCFs in the
time domain has a direct relation to the  Fourier-frequency-dependent  time
lags.

\subsection{Autocorrelation Functions}
\label{sect:acf}

\begin{figure*}
\centerline{\epsfxsize=9.3cm \epsfysize=8.0cm \epsfbox{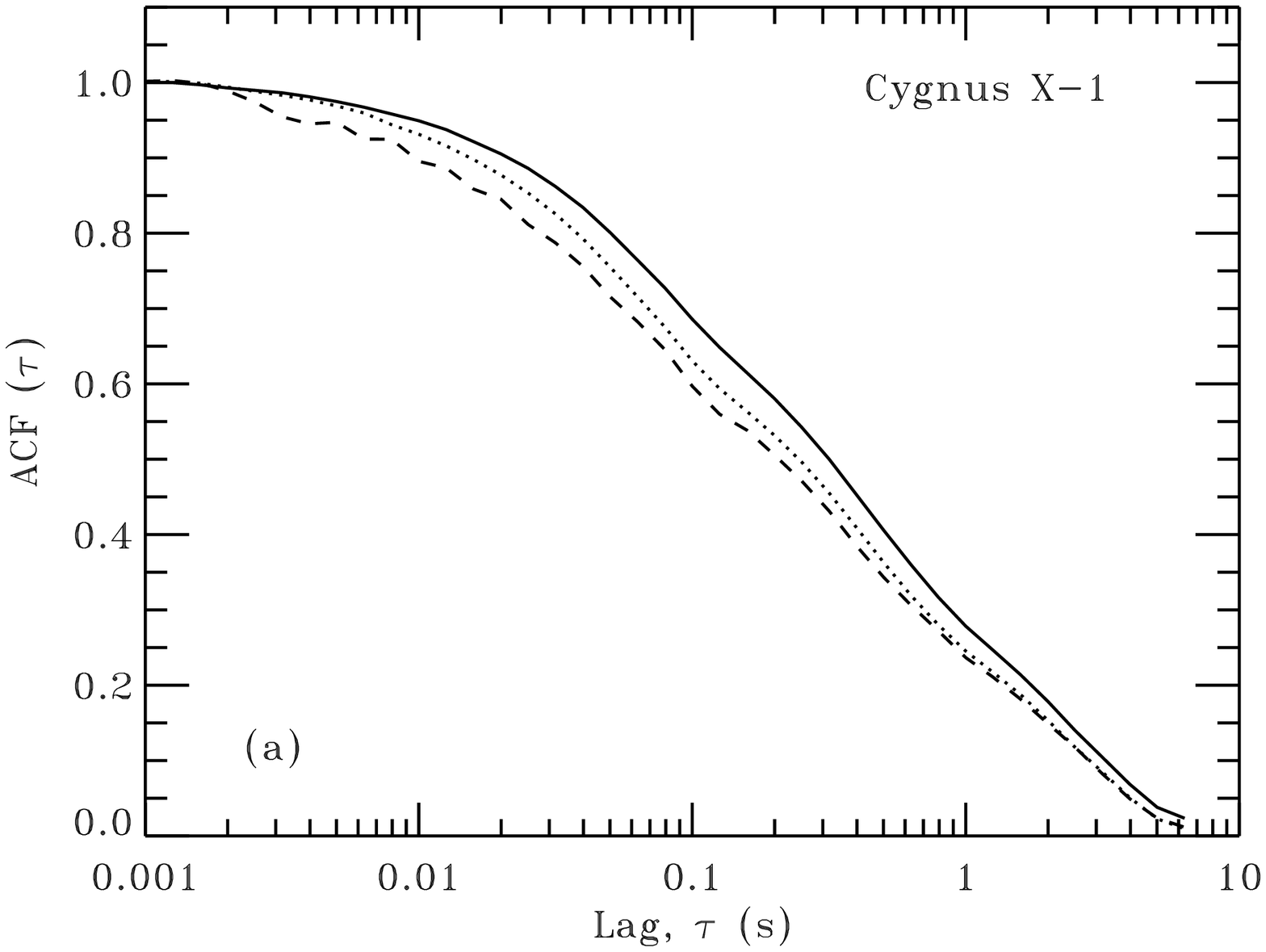}
\epsfxsize=9.3cm  \epsfysize=8.0cm \epsfbox{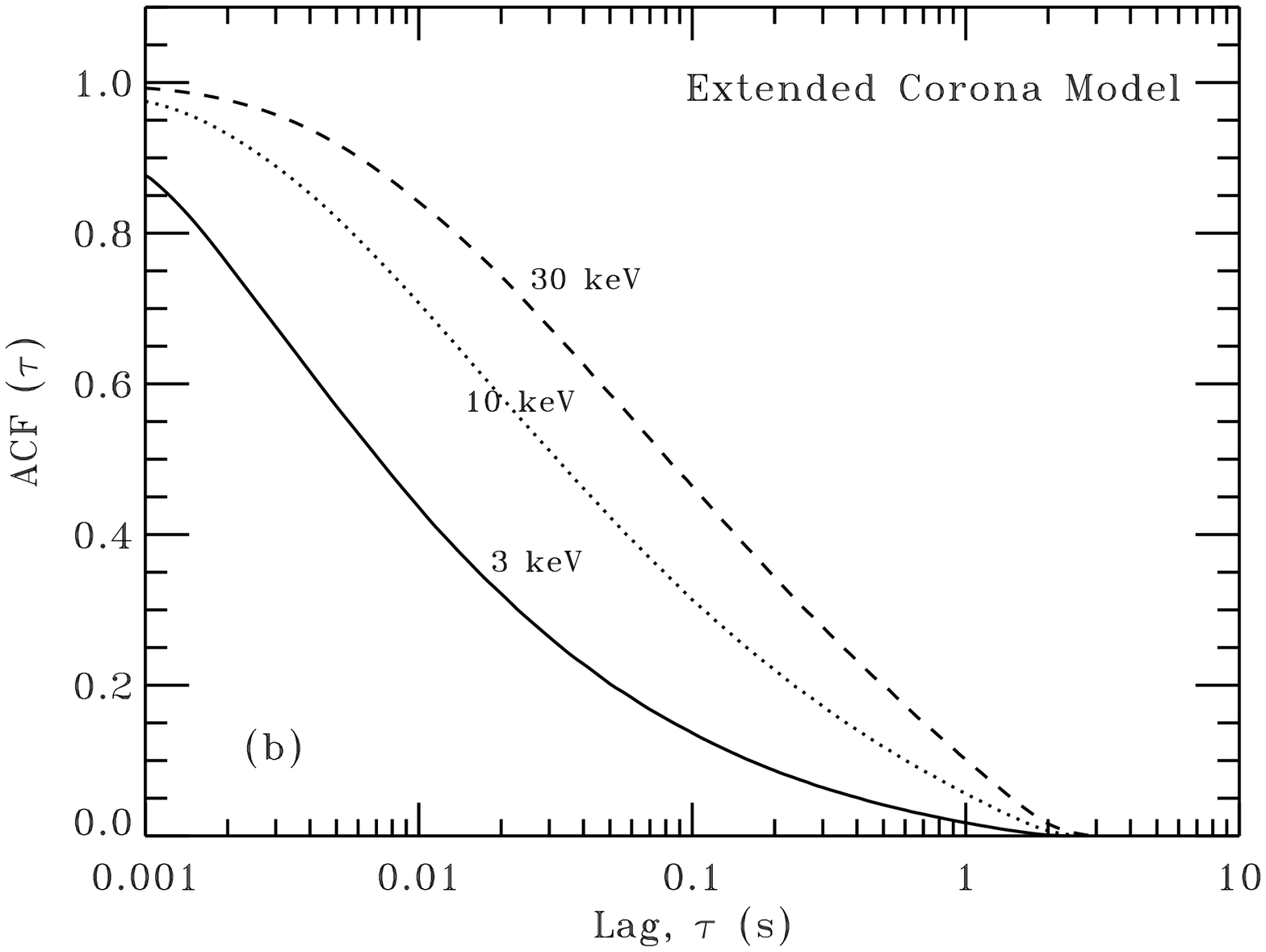}}
%\figcaption{
\caption{ 
The ACFs in the  different  energy  bands.  (a) The data  for  Cygnus  X-1.
Solid curves shows the ACF for the 2-5 keV energy band, dotted curve is for
8-13 keV band, and dashed  for 24-40 keV band.  (b) The  prediction  of the
extended corona model.
\label{fig:acf}}
\end{figure*}

The   autocorrelation   functions   (ACFs)  of  Cygnus  X-1  are  shown  in
Figure~\ref{fig:ccf}  and  Figure~\ref{fig:acf}a.  The  width  of  the  ACF
decreases with photon energy  approximately as $\propto E^{-0.21 \pm 0.01}$
at lags smaller than $\sim 0.3$ sec.  (At larger lags the ACFs at different
energies are not self-similar.)  This strongly constrains the origin of the
spectral  variability,  since it  requires  that the pulses  producing  the
variability  last longer at low energies than at higher  energies.  Similar
energy  dependence is observed also in the ACF of the peak aligned  average
shot  profiles  (Feng  et al.  1999).  Our  results  extend  their  work by
demonstrating  the trend of width of the ACF versus energy  applies  across
more energy  bands.  More  importantly,  we prove that this trend is always
present  in the data, and is not  subject  to the  selection  effects  of a
shot-fitting algorithm.

%\bigskip
%%%%%%%%%%%%%%%%%%%%%%%%%%%%%%%%%%%%%%%%%%%%%

%\bigskip

%\section{Comparison to Model Predictions}

\section{Constraints on the Emission Models}

\subsection{Implications on Shot Shapes for Shot Noise Models}

One can attempt to explain the observed ACFs and CCFs in terms of a
simple shot noise model (Terrell 1972).  Since the CCFs peak at a lag
$\lesssim 2$ ms, the shots at different energies also should reach
maxima within 2 ms from each other.  We can develop some intuition
about the typical shot profiles by looking at the analytic form of the
CCF for shots which have exponential rise and decay and peak at the
same time for different energies.  The CCF is
\begin{equation}
\propto \frac{\tsd^2(\thr+\thd)e^{\tau/\tsd}}{(\thd+\tsd)(\tsd-\thr)}  
- \frac{\thr^2 (\tsr+\tsd) e^{\tau/\thr}}{(\thr+\tsr)(\tsd-\thr)} 
\end{equation}   
for   negative    $\tau$,   and 
\begin{equation} 
\propto   \frac{\thd^2 (\tsr+\tsd) e^{-\tau/\thd}}{(\tsd+\thd)(\thd-\tsr)}
-\frac{\tsr^2 (\thr+\thd)e^{-\tau/\tsr}}{(\tsr+\thr)(\thd-\tsr)} 
\end{equation}   
for positive  $\tau$, where $\tsr,\tsd$  and $\thr,\thd$ are  the rise
and decay time  constants of the shots in soft  and hard energy bands,
respectively.      The    proportionality     coefficient     is    $2
/\sqrt{(\tsr+\tsd)  (\thr+\thd)}$.  The energy  dependence of  the ACF
requires   the  shots  at   higher  energies   to  be   shorter,  i.e.
$\thd+\thr<\tsr+\tsd$.  Since the observed CCF (see Fig.~1) is roughly
equal to the ACF of the softer energy band for positive $\tau$ and the
ACF of the  harder energy band for negative $\tau$,  and the CCF rises
faster    than   it    decays,   one    gets   $\max(\tsd,\thr)\approx
\max(\thr,\thd)  <  \max(\tsr,\thd)  \approx  \max(\tsr,\tsd)$.   This
means that  the soft  rise time  scale is the  longest, $\tsr  > \tsd,
\thr, \thd$, and  that the decay times at  different energies are very
close to one  another (i.e., $\tsd\approx \thd$) or  are so small that
the  only relevant  parameters are  the  rise times  ($\thr \gg  \tsd,
\thd$).  The energy dependence of the rise time then produces the time
lags and the asymmetry of the  CCF (see also Miyamoto \& Kitamoto 1989
who arrived at similar conclusions).

%\subsection{Modified Shot Noise Models}

In modified shot noise models (see, e.g., Lochner et al.  1991), there
is a broad distribution of shot time scales.  The shots at different
energies should be perfectly synchronized in order to achieve high
values of the CCFs.  However, constraints on the shape of the shots
are not so strong as for the simplest shot noise model.  The shots at
different energies can have similar properties as described above, or
they can be shifted relative to each other (e.g., Poutanen \& Fabian
1999a,b).  If the shift depends on the shot time scale as $\propto
\tau^{\alpha}$, one obtains Fourier time lags $\delta t(f)\propto
f^{-\alpha}$.  In such a case, the CCF peaks at a lag equal to the
delay corresponding to the {\it shortest} time scale.  The data then
constrain the minimum shot time scale to be $\lesssim 1$ ms.

\subsection{Extended Corona Models}

Comptonization in  a uniform electron  cloud produces time  lags which
are frequency independent in the range we can probe with current X-Ray
instruments (Miyamoto  et al.  1988).  This inspired  Kazanas, Hua, \&
Titarchuk (1997) to  propose an extended corona model  with a $r^{-1}$
radial density  distribution and a size  of a few  light seconds.  The
long lags  observed at  low Fourier frequencies  are produced  here by
photons travelling and scattering  over large radii, while the shorter
time lags  are produced in the  central small core of  the cloud.  The
model fits  much of the data,  but has the physical  problem of having
too much energy input at large radii.

We simulated the light curves for this model and computed the
auto-correlation functions at different energies.  As an example, we
consider a set of parameters from Hua, Kazanas, \& Cui (1999) ($kT_r=
0.2 \keV$, $kT_e=100\keV$, $p=1$, $n_i=n_1=10^{16}\cm^{-3}$,
$r_1=10^{-3}$~lt-s, $r_2=10^3 r_1$, for details see discussion around
their Fig.~1).  The shots are assumed to be produced by modulations in
the soft flux.  Because this model produces delays by having larger
light travel times for higher energy photons, it always produces a
wider shot and hence a wider ACF at higher energies (see
Fig.~\ref{fig:acf}b).  We note that the ACF at 30 keV in this model is
a factor of ten {\it broader} than the ACF at 3 keV, while in the
observations it is 50\% {\it narrower}.  The fact that the ACFs become
broader with energy is the intrinsic property of the model and there
is no way to resolve this problem by changing the parameters of the
system.  We conclude that models where the time lags are produced by
light travel delays can be ruled out.

\subsection{Magnetic Flare Models}

Magnetic flares on the surface of the cold accretion disk were shown
recently to produce X/$\gamma$-ray spectra in agreement with
observations of Cyg X-1 (Beloborodov 1999b).  The observed time lags
can correspond to the time scale of the evolution of magnetic
structures (Poutanen \& Fabian 1999a,b).  The magnetic field lines
twist due to differential rotation in the accretion disk and elevate
to the corona releasing magnetic energy and heating the corona.  The
flare time scale is then of the order of the Keplerian time scale at
the relevant distance from the central black hole.  Changes in the
energy dissipation rate and in the geometry of the flare (e.g.,
distance from the disk) produce soft-to-hard spectral evolution which
is the cause of the hard time lags.  A broad distribution of the flare
time scale assures that the time lags are inversely proportional to
the Fourier frequency $\delta t(f)\propto \tau_f \sim 1/(2\pi f)$,
where $\tau_f$ is the shot time scale giving contribution to the power
spectrum at frequency $f$.

The model of Poutanen \& Fabian (1999b) reproduces well the time lags
observed in Cyg X-1, while producing a somewhat wider ACF at higher
energies due to the assumption that the energy dissipation rate rises
faster than it decays.  In the opposite situation, when the
dissipation rate rises more slowly than it decays, the ACF and CCF
energy dependences can be reproduced easily (see, e.g., Miyamoto \&
Kitamoto 1989; Poutanen 2000).

\subsection{Drifting Blob Models}

B\"ottcher  \&  Liang  (1999)  proposed  a  model  in  which  spectral
variability  is produced  by a  cool blob  drifting inward  through an
inhomogeneous hot  inner disk.   This model qualitatively  matches the
data.   However,  the  parameters  presented  in  that  work,  produce
significant quantitative deviations from  the observations in both the
Fourier and time domains.  While changing the parameters of the system
could allow the model to fit the data, a larger problem for this model
is  that it  drives the  variability through  modulations of  the soft
photon flux.   If the energy dissipation  rate in the  corona does not
change with time, then an increase  of the soft photon flux would lead
to a softer spectrum and  spectral pivoting around $\sim$ 10 keV.  The
amplitude of the variability would be then a strong function of photon
energy (larger at larger energies) and the variability above and below
the pivoting point would  anti-correlate, contrary to what is observed
(see, e.g., Nowak  et al.  1999a and \S~\ref{sect:ccf}).   In order to
reproduce  the observed correlated  variability at  different energies
without  violating the  energy balance,  one  has to  assume that  the
inward drift of  cool blobs is perfectly correlated  with the increase
of the  energy dissipation in the  hot corona.  It remains  to be seen
whether such a  requirement is physically realistic and  the model can
indeed fit the data.  One cannot avoid this constraint by proposing an
ADAF-type  solution where  the coronal  cooling rate  is  dominated by
cyclo-/synchrotron radiation  rather than by Compton  cooling.  Such a
model  would  require  a  cyclo-/synchrotron luminosity  (observed  in
IR/optical  light) far  in excess  of  the X-Ray  luminosity of  $\sim
10^{37}$ ergs/sec.   The observed optical  luminosity is only  about $5
\times 10^{36}$ ergs/sec and is dominated by the companion star.

\subsection{Cylindrical Wave Models}

Cylindrical  waves propagating  through  the accretion  disk from  the
region where the soft X-rays are  emitted to the region where the hard
X-rays are emitted  (e.g., Miyamoto et al.  1988;  Kato 1989; Nowak et
al.  1999b) have also been  proposed as a mechanism for producing hard
time lags.  The dispersion of wave velocities results in the frequency
dependence of the  time lags.  A recent and  relatively well developed
version  of such  a model  (a ``transition  disk'' model)  designed to
explained  the spectrum  of  Cyg X-1  and  the Fourier  time lags  was
recently considered by Misra (2000).   However, in this model the peak
of the emission at  30 keV is delayed by $\sim 0.015$  s from the peak
at  3 keV.  In  such a  case, the  CCF between  the 30  keV and  3 keV
photons would peak  at that lag strongly contradicting  the data.  One
can assume  that the low frequency signal  propagates slowly producing
larger  time  lags,  while  high frequency  signal  propagates  faster
producing smaller time lags.  (This is basically a modified shot noise
model.)  The  transition disk model then requires  a propagation speed
of $\gtrsim 2 c$ in order to fit $\lesssim 2$ ms lag in the CCF.

\section{Conclusions}

We present the results of cross-correlation and autocorrelation
analysis of the light curve of Cygnus X-1 in several energy bands.
The width of the ACF was found to become narrower with photon energy
as $\propto E^{-0.2}$.  The corresponding CCFs are asymmetric, but all
peak at lags less than 2 ms.

We compare these results to model calculations for an extended Compton
corona model,  a magnetic  flare model, a  drifting blob model,  and a
cylindrical wave model.  The extended corona model inherently produces
longer   shots   at   higher   energies   (and,   therefore,   broader
autocorrelation   function),  and   it  cannot   fit  the   data  even
qualitatively.  We  find that a magnetic  flare model can  fit all the
data if  one requires  that the energy  dissipation rate  rises slower
than it decays.  We also find  that while the drifting blob model fits
the data qualitatively, it requires  the inward drift of cool blobs to
be perfectly correlated with the increase of the energy dissipation in
the hot corona.  A transition disk model predicts a large shift of the
peak of the cross-correlation function from zero lag contradicting the
data.

\acknowledgments

We thank Steve  Kahn for useful discussion and  Andrei Beloborodov for
valuable  comments.  This research  has been  supported by  NASA grant
{\bf  \# NAGS-6691  and  NAGS-7409}  (TJM, PSC),  and  by the  Swedish
Natural  Science  Research  Council  and  the  Anna-Greta  and  Holger
Crafoord Fund (JP).

\end{document}